\documentclass[twocolumn,prl]{revtex4}
\usepackage{epsfig}
\usepackage{graphicx}
\usepackage{psfig,epsfig}
\usepackage{dcolumn}
\usepackage{bm}

\def\e{\mbox{e}}

\begin{document}

\title{Statistical Mechanics of segregation in binary granular mixtures} 
\author{M. Tarzia$^{a}$, A. Fierro $^{a,b}$, M. Nicodemi 
$^{a,b}$, M. Pica Ciamarra $^{a}$, A. Coniglio $^{a,b}$}
\affiliation{${}^a$ Dip. di Scienze Fisiche, Universit\`{a} degli Studi
di Napoli ``Federico II'', INFM and INFN, via Cintia, Napoli, Italy}
\affiliation{${}^b$ INFM - Coherentia, Napoli, Italy} 

\date{\today}

\begin{abstract}
We discuss mixing/segregation phenomena in a schematic hard spheres 
lattice model for binary mixtures of granular media, by analytical evaluation, 
within Bethe-Peierls approximation, of Edwards' partition function. 
The presence of fluid-crystal phase transitions in the system drives 
segregation as a form of phase separation. 
Within a pure phase, gravity can also induce a kind of vertical 
segregation not associated to phase transitions. 
\end{abstract}

\maketitle

Granular mixtures subject to shaking can mix or, under different conditions, 
spontaneously segregate their components according to criteria which, 
although of deep practical and conceptual importance, are still largely 
unclear \cite{rev_segr,capri}. 
In this respect, microscopical grains properties are known to play a role 
since, for example, smaller grains appear to filter beneath larger ones 
(the so called ``percolation'' 
phenomena \cite{rosato,bridgewater}). 
``Inertia'' \cite{shinbrot}, ``convection'' \cite{knight} and 
``dynamical'' effects in general are also known to be relevant \cite{jenkins}.
Interestingly, recent simulations and experiments have outlined 
that segregation processes can also involve ``global'' mechanisms: 
phenomena such as ``condensation'' \cite{luding} or, more generally, 
``phase separation'' \cite{kakalios} have been found, recalling well known 
properties in the Statistical Mechanics of thermal fluids, such as 
colloidal mixtures (see Ref.s in \cite{kakalios,sollich}). 
Actually, even though granular media are not characterized by Boltzmann 
distributions \cite{ng}, fluidized mixtures have been schematically treated 
as standard fluids under gravity in presence of a finite bath 
temperature \cite{luding,both}. 

A different Statistical Mechanics approach was introduced by Edwards 
\cite{Edwards1,nfc,e1} (see \cite{capri} for a review). This was explicitly 
designed to deal with non-thermal systems, such as granular media in their 
``jammed states'' (i.e., not in their fluidized regime): time averages 
of a granular system (subject to some drive, e.g., tapping) 
observed at rest, are supposed to coincide with suitable ensemble averages 
over its ``mechanically stable'' states, i.e., those where the system is found 
still. Even though the limit of validity of such an approach must be 
still assessed, it appears to be well grounded in several cases 
\cite{Edwards1,nfc,e1,capri}. In particular, it was shown to hold 
in a schematic model of granular mixtures:
a lattice binary hard spheres system under gravity subject to sequences 
of taps \cite{nfc}. 
In this paper we analytically solve, at the level of Bethe approximation, 
the partition function \'a la Edwards of such a model and derive its 
phase diagram and mixing/segregation properties as a function of grains 
masses, sizes, number and, see below, ``configurational temperatures''. 

Our model exhibits phase transitions from fluid to crystal phases. 
As much as in thermal media, this induces segregation effects associated 
to the presence of phase separation phenomena, 
with the formation of coexisting phases rich in small or large grains. 
Gravity drives vertical segregation, where large grains can be found 
on average above (the well known ``Brazil nut'' 
effect, BNE \cite{rev_segr}) or below small grains (reverse BNE, RBNE 
\cite{luding,breu}). When associated to phase separation, we have a form of 
{\em strong segregation}, as opposed to {\em mixing}, or similarly 
weak segregation, observed within a given phase. 
Coarsening phenomena, as those experimentally observed \cite{kakalios}, 
are predicted to occur for suitable values of the system parameters. 

The model under investigation \cite{nfc} is a binary hard sphere mixture 
made of two species, 1 (small) and 2 (large) with grain diameters $a_0=1$ 
and $\sqrt{2} a_0$, under gravity on a cubic lattice, of spacing $a_0$, 
confined in a rigid box. 
On each lattice site we define an occupancy variable: $n_i^z=0,1,2$ if 
site $i$ at height $z$ is empty, filled by a small or by a large grain.
The system Hamiltonian is: 
${\cal H}={\cal H}_{HC}+ m_1gH_1 + m_2gH_2$, 
where $H_1=\sum_{i,z} z \, \delta_{n_i^z 1}$, 
$H_2=\sum_{i,z} z \,\delta_{n_i^z 2} $ are the heights of the two species 
and ${\cal H}_{HC}$ is the hard core potential, preventing two nearest 
neighbor sites to be both occupied if at least one contains a large grain. 

\begin{figure}[ht]
\begin{center}
\hspace{4.6cm}\vspace{-5cm}\psfig{figure=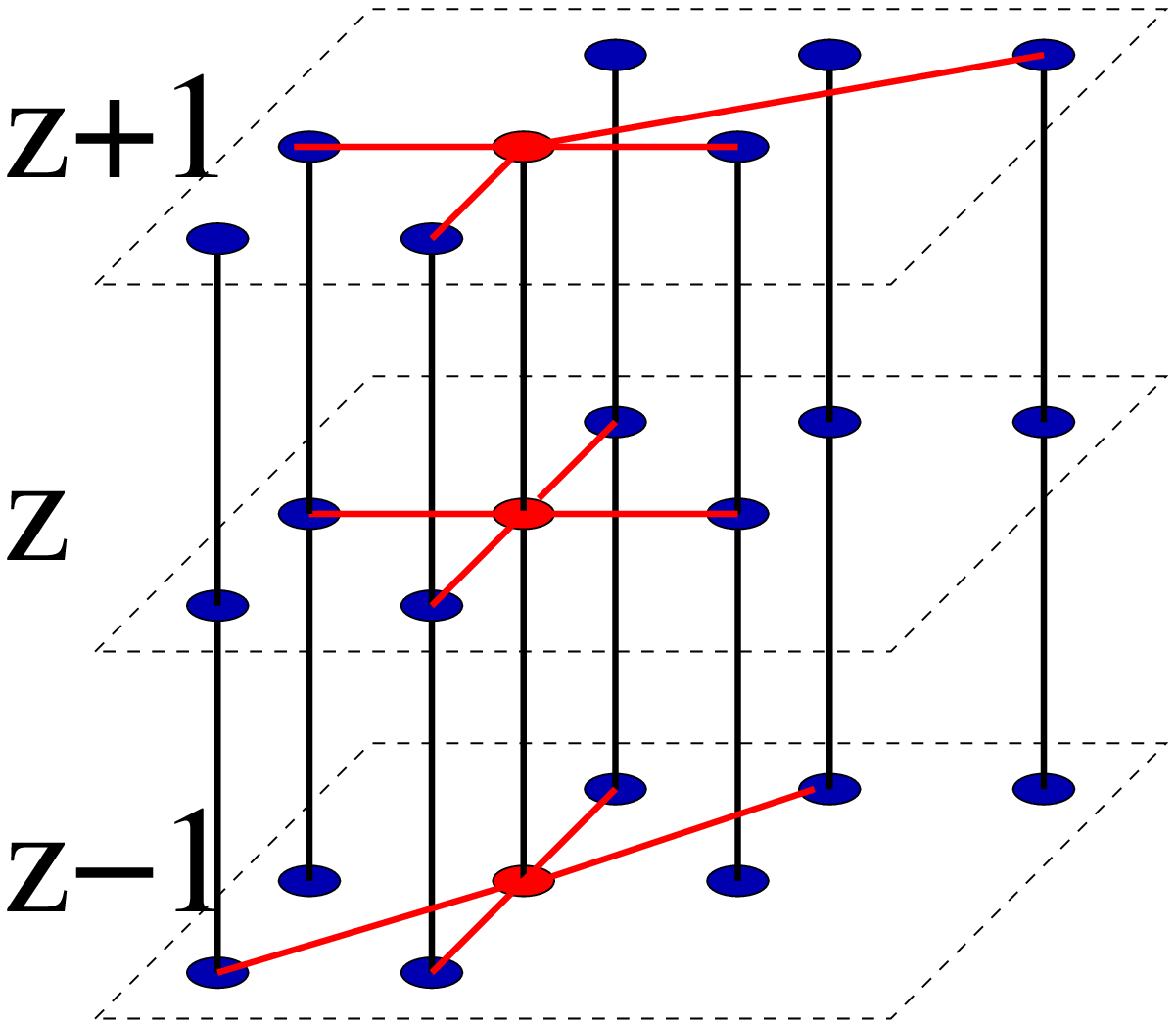,width=3.2cm,angle=0}
 
\hspace{-2cm}\psfig{figure=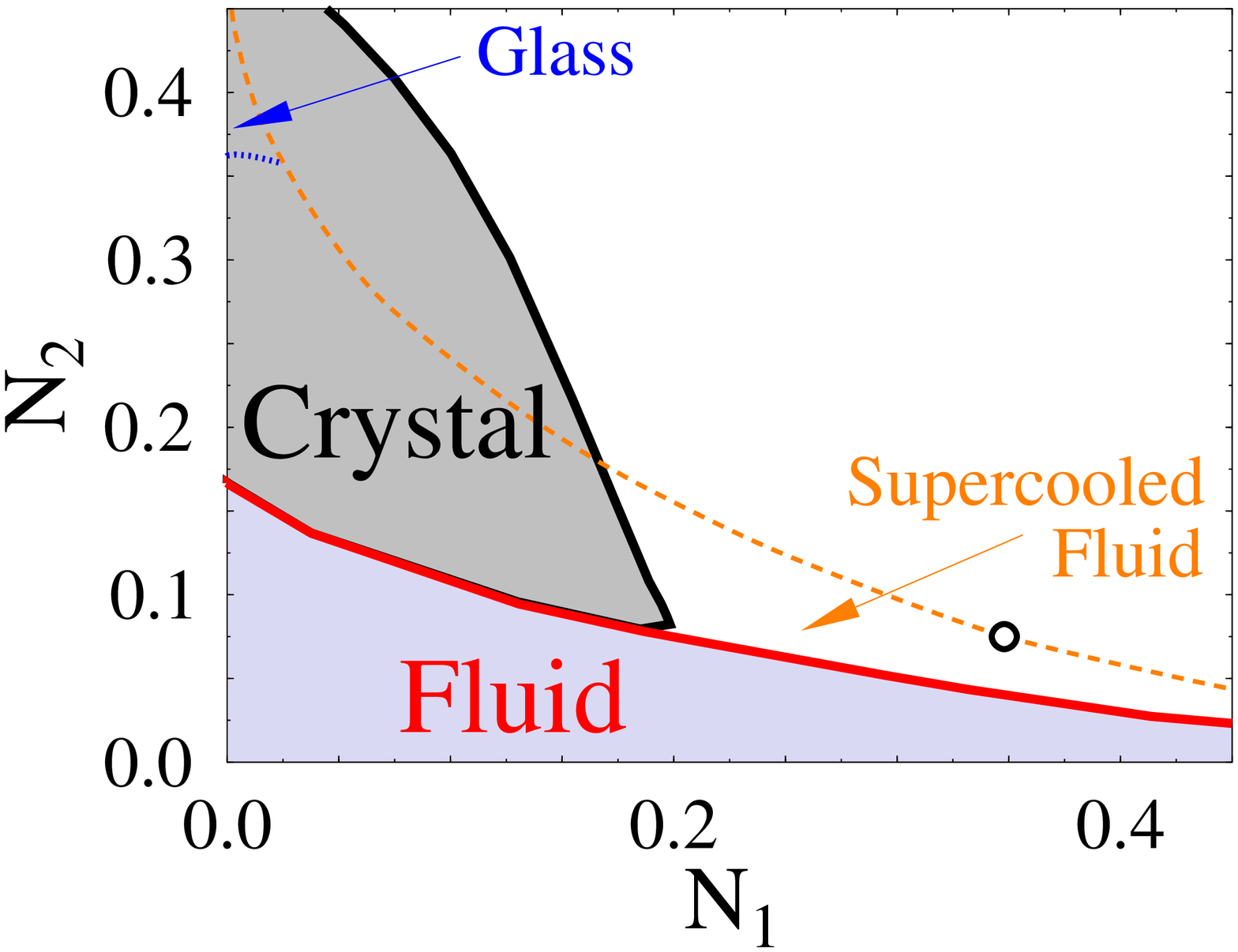,width=10cm,angle=-90}
\end{center}
\vspace{-2.2cm}
\caption{{\bf Main frame} 
Phase diagram, from Bethe approximation, of the hard sphere binary mixture 
model confined to a single 2D horizontal layer. Here 
the control parameters are the densities per unit surface 
of small and large grains $N_{1}$, $N_{2}$. 
A Fluid and a Crystal phase are found, 
divided by a discontinuous melting transition. 
When the mixture is prepared with values of $N_{1}$ and $N_{2}$ 
out of the Fluid or Crystal phases (i.e., in the white area), 
it exhibits phase separation, associated to coarsening, in a fluid phase 
rich in small grains and a crystalline phase rich in large ones. 
This causes horizontal segregation of the two grains species. 
We also plot the metastable ``supercooled'' fluid region, where a 
coexistence line between two supercooled fluids, with different relative 
species concentrations, is found (dashed line) having a critical 
point (open dot). 
For high $N_{2}$, a 1-step Replica Symmetry Breaking glassy phase 
(described in details elsewhere \cite{cdfnt}), metastable with respect 
to the crystal, is present. 
{\bf Inset} A schematic picture of the 3D lattice model: we consider a 
binary mixture of hard spheres located on a Bethe lattice 
where each horizontal layer is a random graph of given connectivity. 
Homologous sites on neighboring layers are also linked and the overall 
vertex connectivity is $c\equiv k+1=6$.
} \label{piano}
\end{figure}

As stated, Edwards' approach appears to hold with good approximation 
in this model \cite{nfc} under a tap dynamics (analogous to that used in the
compaction of real granular
media), which turns out to be characterized by two 
suitable ``thermodynamic parameters'', 
conjugated to the heights of the two species and 
called configurational temperatures, $T^{conf}_1$ and $T^{conf}_2$.
Their values are increasing functions of the tap amplitude, 
$\Gamma$, and tend to coincide for $\Gamma\rightarrow 0$ \cite{prep}. 
In brief, the weight of a given microstate $r$ turns out to be \cite{nfc}:
$\exp\left\{-{\cal H}_{HC}(r)-\beta_1m_1gH_1(r) -\beta_2m_2gH_2(r)\right\}\cdot \Pi_r$,
where $T^{conf}_1 \equiv \beta_1^{-1}$ and $T^{conf}_2 \equiv \beta_2^{-1}$. 
The operator $\Pi_r$ selects mechanically stable states:
$\Pi_r=1$ if $r$ is ``stable'', else $\Pi_r=0$.
The system partition function \'a la Edwards is thus the following \cite{nfc}:
\begin{equation}
\nonumber
{\cal Z} =  \sum_{\{r\}} \e^{-\left[{\cal H}_{HC} (r) + \beta_1 
m_1gH_1 
+\beta_2 
m_2gH_2 
\right]}
\cdot \Pi_r
\label{Zed}
\end{equation}
where the sum is over all microstates $r$. 
Since the exact calculation of ${\cal Z}$ is hardly feasible, 
we want to evaluate ${\cal Z}$ here at a mean field level. To this aim 
we consider a generalization of Bethe-Peierls method for anisotropic systems 
(due to gravity), i.e., we solve by recurrence relations the 
partition function of the system on the Bethe lattice shown in 
the inset of Fig.\ref{piano} (see also \cite{MP,cdfnt}). 
The reader not interested in technicalities about calculations can jump 
to the paragraph after Eq.(\ref{ricorrenza}) where the results are discussed. 
In particular, we consider 
a 3D lattice box with $H$ horizontal layers (i.e., $z\in\{1,...,H\}$) 
occupied by hard spheres. Each layer is a random graph of given 
connectivity, $k-1$ (we take $k=5$). Each site in layer $z$ is also
connected to its homologous site in $z-1$ and $z+1$
(the total connectivity is thus $k+1$).
The Hamiltonian is the one above 
plus two chemical potential terms to control the two species densities 
$(\mu_1,\mu_2)$. 
Hard Core repulsion prevents two particles on connected sites to overlap. 
In the present lattice model we adopt a simple definition of ``mechanical
stability'': a grain is ``stable'' if it has a grain underneath.
For a given grains configuration 
$r=\{n_i\}$, the operator $\Pi_r$ has a tractable expression:
$\Pi_r =\lim_{K\rightarrow\infty}\exp\left\{-K{\cal H}_{Edw}\right\}$
where ${\cal H}_{Edw}= 
\sum_{i,z} \left[ \delta_{n_i^z 2}\delta_{n_i^{z-1} 0}\delta_{n_i^{z-2} 0} 
+ \delta_{n_i^z 1}\delta_{n_i^{z-1} 0}
\left(1 - \delta_{n_i^{z-2} 2}\right) \right]$.

The local tree-like properties of our lattice allows to write down recursive 
equations \'a la Bethe found by iteration of the lattice structure. 
Details on calculations and notations can be found in \cite{capri,MP,cdfnt}, 
here we just recall the main lines. 
A ``branch'' of our lattice can 
iteratively grow in the ``up'', ``down'' or  ``side'' directions. 
The partition functions of new branches starting from 
a site $i$ at height $z$ can be 
recursively written in terms of their old ones. To this aim 
define $Z_{0,s}^{(i,z)}$ and $Z_{n,s}^{(i,z)}$ the partition functions 
of the ``side'' branch restricted respectively to configurations 
in which the site $i$ is empty or filled by a particle of specie $n$
(with $n=1,2$); analogously, $Z_{n,u}^{(i,z)}$, 
$Z_{0,u}^{(i,z)}$ and $\overline{Z}_{n,u}^{(i,z)}$ 
(resp. $Z_{n,d}^{(i,z)}$, $Z_{0,d}^{(i,z)}$
and $\overline{Z}_{n,d}^{(i,z)}$)
are the partition functions of the ``up'' (resp. ``down'') 
branch restricted to 
configurations in which the site $i$ is filled by a 
grain of specie $n$, empty with the upper (resp. lower) site 
empty and empty with the upper (resp. lower) site filled by a 
grain of specie $n$. 
In order to write Bethe recursive equations, 
it is convenient to introduce the local ``cavity fields'' defined by: 
$e^{s_{n}^{(i,z)}} = Z_{n,s}^{(i,z)}/Z_{0,s}^{(i,z)}$,
$e^{u_{n}^{(i,z)}} = Z_{n,u}^{(i,z)}/Z_{0,u}^{(i,z)}$,
$e^{v_{n}^{(i,z)}} = \overline{Z}_{n,u}^{(i,z)}/Z_{0,u}^{(i,z)}$,
$e^{d_{n}^{(i,z)}} = Z_{n,d}^{(i,z)}/Z_{0,d}^{(i,z)}$,
$e^{c_{n}^{(i,z)}} = \overline{Z}_{n,d}^{(i,z)}/Z_{0,d}^{(i,z)}$
(with $n=1,2$).
In these new variables the recursion relations are more easily written:
\begin{eqnarray} \label{ricorrenza}
\nonumber
e^{s_{n}^{(i,z)}} & = & \frac{e^{\beta_n(\mu_n-m_ngz)}}
{Q^{(i,z)}} \left[
\prod_{j=1}^{k-2}
\frac{A_n^{(j,z)}}{S^{(j,z)}} \right]
T_n^{(i,z+1)} R_n^{(i,z-1)}
\\
\nonumber
e^{u_{n}^{(i,z)}} & = & e^{\beta_n(\mu_n-m_ngz)}
\left[\prod_{j=1}^{k-1}
\frac{A_n^{(j,z)}}{S^{(j,z)}}\right] T_n^{(i,z+1)}
\\
e^{v_{n}^{(i,z)}}&=&e^{u_{n}^{(i,z+1)}} 
\\
\nonumber
e^{d_{n}^{(i,z)}} & = & e^{\beta_n(\mu_n-m_ngz)}
\left[\prod_{j=1}^{k-1}
\frac{A_n^{(j,z)}}{S^{(j,z)}}\right]R_n^{(i,z-1)}
\\
\nonumber
e^{c_{n}^{(i,z)}} & = & e^{d_{n}^{(i,z-1)}} \left[1 + R_2^{(i,z-1)}
\right]^{-1}
\end{eqnarray}
Here $n=1,2$, the products run on the neighbors, $j$, of site $i$, 
$A_n^{(j,z)} = 1+\delta_{n,1}e^{s_{1}^{(j,z)}}$, 
$T_1^{(j,z)} = 1+ e^{u_{1}^{(j,z)}} + e^{v_{2}^{(j,z)}}$,
$T_2^{(j,z)} = 1+ e^{v_{2,u}^{(j,z)}} + e^{v_{1,u}^{(j,z)}}$,
$R_1^{(j,z)} = e^{d_{1}^{(j,z)}}
+ e^{c_{2}^{(j,z)}}$, $R_2^{(j,z)} = e^{c_{2}^{(j,z-1)}}
+ e^{c_{1}^{(j,z-1)}}$,
$S^{(j,z)}=A_1^{(j,z)}+e^{s_{2}^{(j,z)}}$, 
$Q^{(j,z)} = 1+e^{d_{1}^{(j,z-1)}} 
[1+ e^{u_{2}^{(j,z+1)}}]
+e^{d_{2}^{(j,z-1)}} 
[1+e^{u_{1}^{(j,z+1)}} +  e^{u_{2}^{(j,z+1)}}] 
+ R_2^{(j,z)}$.

\begin{figure}[ht]
\vspace{-1.5cm}\centerline{
\hspace{-2.5cm}\psfig{figure=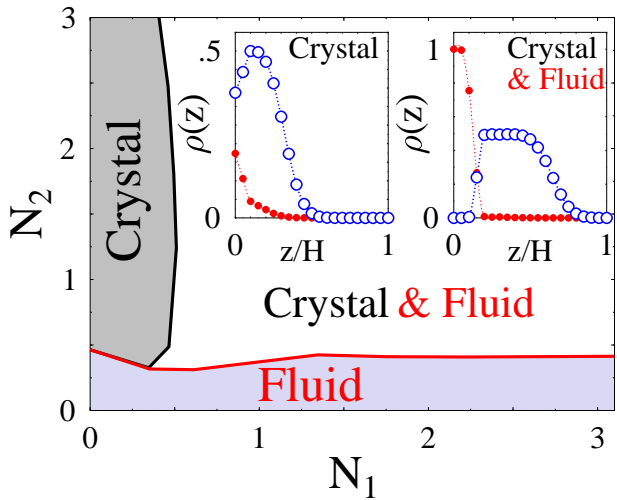,width=10cm,angle=-90}}
\vspace{-1.8cm}
\caption{Phase diagram of a 3D system under gravity, treated \'a la Edwards, 
in the plane $(N_{1}, N_{2})$ for $m_1 \beta_1 = 1$ and $m_2 \beta_2 = 1/2$ 
({\bf main frame}). 
A pure Fluid and Crystal phase are present along with a region where 
a form of phase separation occurs and the 
two phases coexist (the white area marked Crystal\&Fluid): 
since gravity breaks the vertical 
symmetry, here a crystal rich in large grains is resting on a fluid bed rich 
in small grains. This is a strong BNE situation, visualized in the 
{\bf right inset} showing the density profiles of the two species, 
$\rho_1 (z)$ and $\rho_2 (z)$ (resp. filled and empty circles), 
in a typical point of the Crystal\&Fluid region with $N_1=3$, $N_2=4$. 
The reverse, i.e., strong RBNE with the fluid floating above the crystal, 
can be found when $m_1 \beta_1 < m_2 \beta_2$. 
When the configurational temperatures and species densities get close 
coarsening phenomena may occur. 
For comparison in the {\bf left inset} we show $\rho_1 (z)$ and $\rho_2 (z)$ 
as a function of the vertical coordinate $z$, 
in a point of the Crystal phase with $N_1=0.3$, $N_2=4$: small grains are 
here interspersed with large ones even though, on average, slightly below. 
This illustrates that within a pure phase, gravity also drives a weaker 
form of segregation, not associated to phase separation, as discussed 
in the text. 
For clarity, metastable phases are not shown in this phase diagram. 
} \label{2pd}
\end{figure}

From local fields the free energy, $F$, can be derived \cite{capri,MP,cdfnt} 
along with the quantities of interest, such as the 
density profile of small and large grains, $\rho_1 (z)$ and $\rho_2 (z)$, 
their density per unit surface 
$N_{1} = \sum_z \rho_1 (z)$ and $N_{2} = \sum_z \rho_2 (z)$ 
and average heights 
$h_n=\langle{z_n}\rangle=\sum_z z\rho_n(z)/\sum_z\rho_n(z)$ (with $n=1,2$). 
The system parameters (for a given grains sizes ratio) 
are four: the two number densities per unit surface, $N_{1}$ and $N_{2}$
(conjugated to the chemical potentials), 
and the two configurational temperatures, or more precisely 
$m_1 \beta_1$ and $m_2 \beta_2$ (conjugated to gravitational energies). 
In the space of these parameters, 
the fluid phase corresponds to a solution of Bethe-Peierls equations 
where local fields in each layer are site independent. 
Such a solution, characterized by horizontal translational invariance, is 
given by the fixed points of Eqs.(\ref{ricorrenza}). 
With the standard Bethe-Peierls methods, a crystalline phase, characterized by
the breakdown of the translational invariance (local fields are now 
different on neighboring sites), is also found. 
The typical resulting phase diagrams are shown in 
Fig.s \ref{piano} and \ref{2pd}. 

For sake of clarity, we first describe the case where the system is 
confined to a 2D horizontal layer (see Fig. \ref{piano}). 
In this case gravity 
plays no role (i.e., we can set $g=0$) 
and the only control parameters are $N_{1}$ and $N_{2}$. 
Our findings are consistent with known results on binary hard spheres 
with no gravity \cite{evans}. 
A Fluid phase is found for small $N_{2}$ and, by increasing $N_{2}$, 
there is a discontinuous (except for $N_{1}=0$) 
melting transition to a Crystal phase. 
Whenever the system is prepared with values of $N_{1}$ and $N_{2}$ 
out of the regions where these 
phases are stable (i.e., in the white area of Fig.\ref{piano}), 
phase separation occurs between a fluid rich in small 
grains and a crystal rich in large grains \cite{note_2ndcryst}. 
This might correspond to a horizontal segregation in the grains mixture, 
as recently found in the form of coarsening phenomena in experiments 
on a monolayer by Reis and Mullin \cite{kakalios}.
Fig. \ref{piano} also shows the ``supercooled fluid'' and ``glassy'' 
\cite{cdfnt} metastable phases (plotted with smaller fonts). 
Since nucleation times can be in practice very long (and enhanced by a 
degree of polidispersity), crystallization can be avoided and the metastable 
fluid observed. In such a region 
we find a coexistence line between a fluid rich
in small grains and another one rich in big grains (dashed line in 
Fig. \ref{piano}) with a second order critical point (big empty circle). 

Fig. \ref{2pd} shows the typical phase diagram of a 3D system in a case 
where $m_1\beta_1>m_2\beta_2$. 
The pure Fluid and Crystal phases found in 2D are still present,  
even though their extension depends now on $\beta_1$ and $\beta_2$ 
(they shrink as $\beta_1$ and $\beta_2$ increase). 
There is also a region where the two phases coexist 
(marked Crystal\&Fluid). 
Gravity breaks the system ``up-down'' symmetry, so the coexisting 
stable or metastable phases of the 2D case are 
vertically segregated. This separation mechanism underlies a form of 
{\em strong vertical segregation} (see Fig.\ref{2pd}): 
here the crystal phase, rich in large grains, moves to the top 
and a clear cut BNE is found (as RBNE is observed in the opposite case, 
when $m_1\beta_1<m_2\beta_2$). 
This kind of segregation is visualized by a plot of the density profiles 
which, in this region, show a clear separation of the two coexisting phases 
(see right inset of Fig.\ref{2pd}). 

Opposed to such phase separation driven segregation, 
within the pure Fluid and Crystal phases 
one observes {\em mixing} or a form of weak vertical segregation since 
more mechanically stable states can be found with small grains at the bottom 
(as described by ``percolation'' mechanisms). 
This is shown, in a typical point of the Crystal phase, by the species 
density profiles plotted in the left inset of Fig.\ref{2pd}: 
small grains are essentially mixed with large ones and, on average, 
slightly below. 
In the phase diagram of Fig.\ref{2pd} different forms of BNE can thus be 
found in different regions 
(similarly for RBNE in the case $m_1 \beta_1 < m_2 \beta_2$). 
In general, for a given grains sizes ratio, 
an interplay of mass densities and configurational 
temperatures difference drives the phases vertical positioning.

Due to the symmetry breaking field, 
in 3D no coarsening phenomena are usually associated to segregation, 
in contrast with the 2D case. Coarsening is expected to appear in the 
segregation process when both the phases densities and the configurational 
temperatures get close, a phenomenon which could be 
confirmed by experiments or simulations. 

\begin{figure}[ht]
\vspace{-1.5cm}\centerline{\hspace{-2.5cm}\psfig{figure=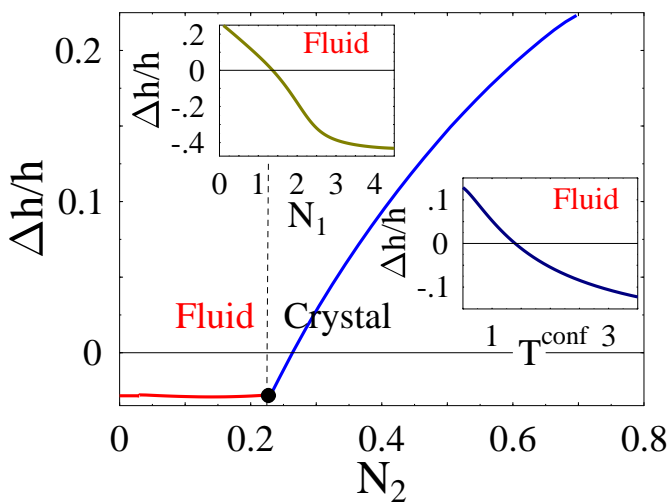,width=10cm,angle=-90}}
\vspace{-2cm}
\caption{
In a 3D system a variety of mixing/segregation behaviors can be observed 
by changing grains and system parameters. {\bf Main frame} 
Given $N_1=0.4$, $m_1 \beta_1=0.8$ and $m_2 \beta_2=1.25$, 
the species relative height difference 
$\Delta h /h$ is plotted as a function of $N_2$ as the system crosses 
from the Fluid to the Crystal phase (see Fig.\ref{2pd}), 
the full dot and the dashed line showing the phase transition point. 
{\bf Upper inset} 
Given $N_2=0.4$, $m_1 \beta_1=1$ and $m_2 \beta_2=2$, 
$\Delta h /h$ is plotted as a function of $N_1$ in the Fluid phase.
{\bf Lower inset} 
Given $N_1=0.5$, $N_2=N_1/2$, $m_1=1$ and $m_2=2$, 
$\Delta h /h$ is plotted as a function of
$T_1^{conf}=T_2^{conf}\equiv T^{conf}$ in the Fluid phase. 
In all these cases $\Delta h /h$ changes sign, passing from a 
BNE to a RBNE region, through a mixing region where 
$\Delta h /h\simeq 0$. 
} \label{dhh3}
\end{figure}

These are the basic mechanisms which underly segregation/mixing phenomena and 
which give rise to a variety of behaviors as grains and system parameters are 
changed. This is illustrated by a few more examples in Fig.\ref{dhh3} selected 
to describe cases where the system moves from weak BNE to weak RBNE regions. 
The picture shows the usual vertical segregation parameter 
$\Delta h /h\equiv 2(h_1-h_2)/(h_1+h_2)$. 
In the main panel of Fig.\ref{dhh3}, $\Delta h /h$ is plotted 
as a function of $N_2$ in a region where the system crosses 
from the fluid to the crystal phase (see Fig.\ref{2pd}), 
the full dot showing the phase transition point: 
well within the crystal, $\Delta h /h$ changes its sign 
passing from BNE, $\Delta h /h<0$, to RBNE, $\Delta h /h>0$.
The insets of Fig.\ref{dhh3} show $\Delta h /h$ in the fluid phase as 
a function of $N_1$ and $T_1^{conf}=T_2^{conf}\equiv T^{conf}$ respectively. 

Summarizing, 
the present mean field Statistical Mechanics model of granular mixture, 
here analytically treated \'a la Edwards, 
allows to explain the basic mechanisms underlying a variety of 
mixing and segregation phenomena experimentally observed, 
ranging from BNE \cite{rev_segr} and RBNE \cite{luding,breu}, 
to coarsening effects \cite{kakalios}. Interestingly for non-thermal media, 
they turn out to be related to thermodynamic-like mechanisms taking place 
in the system phase diagram.

Work supported by MURST-PRIN 2002, MIUR-FIRB 2002, CRdC-AMRA, INFM-PCI.


\begin{thebibliography}{40}
\bibitem{note_2ndcryst} For sake of clarity here we considered the case where 
small grains cannot form a crystalline phase. The latter can be obtained 
with a slightly more complex ${\cal H}_{HC}$.
\end{thebibliography}
\end{document}